\documentstyle[12pt,aps,prc,epsfig,preprint]{revtex}
\tightenlines
\begin{document}
\begin{titlepage}

\title{Saturation of low-energy antiproton annihilation on nuclei}
\author{A. Gal$^a$, E. Friedman$^a$, C.J. Batty$^b$ \\
$^a$Racah Institute of Physics, The Hebrew University, Jerusalem
91904, Israel \\
$^b$Rutherford Appleton Laboratory, Chilton, Didcot, Oxon OX11 0QX, UK}
\maketitle

\begin{abstract}
	Recent measurements of very low-energy ($p_{L}<100$ MeV/c) $\bar p$
        annihilation on light nuclei reveal apparent suppression of annihilation
	upon increasing the atomic charge $Z$ and mass number $A$. Using
	$\bar p$-nucleus optical potentials $V_{{\rm opt}}$, fitted to $\bar
	p$-atom energy-shifts and -widths, we resolve this suppression as due
	to the strong effective repulsion produced by the very absorptive
	$V_{{\rm opt}}$. The low-energy $\bar p$-nucleus
        wavefunction is kept
	substantially outside the nuclear surface and the resulting
	reaction cross section saturates as function of the strength of Im
	$V_{{\rm opt}}$. This feature,
        for $E >0$, parallels the recent prediction,
	for $E < 0$, that the level widths of $\bar p$ atoms saturate and,
	hence, that $\bar p$ deeply bound atomic states are relatively
	narrow. Antiproton annihilation cross sections are calculated at
        $p_{L}=57$ MeV/c across the periodic table, and their dependence
        on $Z$ and $A$ is classified and discussed with respect
        to the Coulomb focussing effect at very low energies.
\newline
$PACS$: 24.10.Ht, 25.43.+t, 25.60.Dz
\newline
 \newline
Corresponding author: A. Gal\newline
Tel: +972 2 658 4930,
Fax: +972 2 561 1519, \newline
E mail: avragal@vms.huji.ac.il
\end{abstract}
\centerline{\today}
\end{titlepage}

\section{Introduction}
\label{sec:int}

Experimental results for antiproton annihilation cross sections
at very low energies ($p_L < 100$ MeV/c),
below the $\bar p p \to \bar n n$ charge-exchange
threshold, have recently been reported for light nuclei
\cite{Zen99a,Zen99b,Bia00a}. At these energies the total $\bar p$
reaction cross section consists only of $\bar p$ annihilation.
Bianconi et al. \cite{Bia00a} reported the first ever measurements of
$\bar p$ annihilation on Ne in the momentum range of 53-63 MeV/c.
Comparing with other data, they showed that
 whereas at relatively
higher energies ($p_{L} \approx 200 - 600$ MeV/c)  ratios of
$\bar p$ annihilation cross sections on different nuclei exhibit the
well-known $A^{2/3}$ strong-absorption dependence, such
ratios at very low energies defy any simple, obvious regularity. It
has been demonstrated that the {\it `expected'} $ZA^{1/3}$ dependence
of these cross sections
on the atomic charge $Z$ and mass number $A$ is badly violated
\cite{Bia00a,Bia00b}. Antiproton annihilation cross sections at very
low energies simply do not rise with $A$ as fast as is anticipated.
For example, over the whole mass range studied so far (H, D,
$^{4}$He, Ne), the Ne/H ratio of total annihilation cross sections
is at least 6 times smaller than expected.

In the present Letter we study low-energy $\bar p$ annihilation on
nuclei, using the optical model approach. Optical potentials have
been very successful in describing strong-interaction effects in
hadronic atoms \cite{BFG97}, including $\bar p$ atoms \cite{BFG95}.
It has been noted, for pions, that the total reaction cross sections
at low energies are directly related to the atomic-state widths, and
that once a suitable optical potential is constructed by reasonably
fitting it to the atomic level shifts and widths in the negative
energy bound-state domain, these total reaction cross sections are
reliably calculable \cite{SMC79,FGJ91}. The recent publications
\cite{Zen99a,Zen99b,Bia00a} of experimental results of total cross
sections for $\bar p$ annihilation on nuclei at very low energies
raise the intriguing possibility of connecting these two energy
regimes in a systematic way also for antiprotons. However, most
of the data on annihilation cross sections for $\bar p$
are for very light nuclei, where the concept of a rather universal
optical potential that depends on $A$ and $Z$  only through the
nuclear densities is questionable. For this reason we use optical
potentials, in the present work, mostly for crossing the $E=0$
borderline within the {\it same} atomic mass $(A)$ range, from bound
states to scattering. This optical model approach is different from
the allegedly model-independent scattering length approximation
\cite{PBL00} which we have found to be  less useful.
We defer a discussion of the latter statement to a separate
publication.

Among the targets used in the $\bar p$ annihilation measurements
\cite{Zen99a,Zen99b,Bia00a}, the only `real' nuclei are $^4$He and Ne,
in the sense that the conventional optical model is not expected
to be applied to the lighter targets of hydrogen or deuterium.
In the present work we show that the very low-energy $\bar p$ total
annihilation cross sections on $^4$He \cite{Zen99b} and on Ne
\cite{Bia00a} are reproduced well by reasonable optical potentials
fitted separately to the $\bar p$ atomic data at the relevant atomic
mass range. These potentials are strongly absorptive, which leads to
a remarkable saturation of the total reaction cross section with
increasing $A$. Strong absorption has very recently been shown
\cite{FGa99a,FGa99b} to lead also to saturation of the widths of
$\bar p$ atomic states and to the prediction of relatively
narrow deeply bound $\bar p$ atomic states. In view of the close
analogy between bound-state widths and total reaction cross sections
(see Eqs. (\ref{equ:gamma},\ref{equ:sigma}) below),
it should come as no surprise that the underlying physics is the same.
In fact, several $\bar p$D calculations \cite{WGN85,LTa90,KPV00} have
found that the $\bar p$ absorptivity in deuterium, as suggested by the
magnitude of the imaginary part of the $\bar p$D $s$-wave scattering
length, is weaker than in hydrogen due partly to the repulsive
effect of absorption.
Lastly, we use $\bar p$ optical potentials fitted to a comprehensive
set of atomic data across the periodic table to predict $\bar p$ total
annihilation cross sections at very low energies for a wide range of
$A$ values, with the exception of very light targets.
We show such calculated cross sections at $p_{L}=57$ MeV/c and
discuss their $A$ and $Z$ dependence,
particularly with respect to the $ZA^{1/3}$ dependence arising from
Coulomb focussing in the limit of very low energies. We also 
demonstrate the extent to which this dependence is violated at 57 MeV/c.

\section{Optical potentials}
\label{sec:oppot}
In the present work we aim at connecting $\bar p$ atoms,
at energies slightly below threshold, with $\bar p$ annihilation on
nuclei at very low energies above threshold, using an optical potential
$V_{\rm opt}$. Assuming for simplicity a Schr\"odinger-type equation,
the width $\Gamma$ of an atomic level is
given (non-perturbatively) by:

\begin{equation} \label{equ:gamma}
\frac{\Gamma}{2}= -\frac{\int {\rm Im} V_{{\rm opt}}(r)
| \psi({\bf r}) | ^2  d {\bf r}}
{\int | \psi({\bf r}) | ^2  d {\bf r}}\quad.
\end{equation}
Here $\psi({\bf r})$ is the $\bar p$ full atomic wavefunction.
The corresponding expression for the total reaction cross section
at positive energies is

\begin{equation} \label{equ:sigma}
\sigma_R = -\frac{2}{\hbar v} \int | \chi({\bf r}) |^2
{\rm Im} V_{{\rm opt}}(r) d {\bf r}\quad.
\end{equation}
Here $\chi({\bf r})$ is the $\bar p$ - nucleus elastic scattering
wavefunction and $v$ is the c.m. velocity.

There exist several optical model potentials \cite{BFG97,BFG95}
parameterized in terms of nuclear densities which are quite
successful in reproducing the strong interaction effects in not too
light nuclei, e.g. fitting all data for nuclei heavier than carbon.
However, most of the good quality data on $\bar p$ annihilation at
very low energies are for very light nuclei where a straightforward
application of the potential deduced from heavier targets is not
expected to yield good fits. We have therefore started the present
analysis by studying $\bar p$ atoms of $^{3,4}$He.

The $\bar p$ nucleus optical potential used here
is given by the `$t\rho$' expression \cite{BFG97}

\begin{equation}\label{equ:potl}
2\mu V_{{\rm opt}}(r) =
 -{4\pi}(1+{\frac{\mu}{m}})b_0
\rho(r) \;\;\; ,
\end{equation}
where $m$ is the nucleon mass, $\mu$ is the reduced $\bar p$-nucleus
mass, $b_0$ is a complex parameter obtained from fits to the data
and $\rho(r)$ is the nuclear density distribution normalized to $A$.
Trying to fit level shift and width data in $^{3,4}$He \cite{Sch91}
for the 2$p$ and 3$d$ states simultaneously, we always ended up
with the calculated width of the 2$p$ state in $^4$He being too small,
thus contributing an unacceptably large value to
the total $\chi ^2$ of the fits. Handling each isotope separately did
not change this situation and we then decided to fit only the
data for the 2$p$ states for the two  isotopes. This is somewhat
unfortunate as one expects to
be able to use $l$-independent potentials, as is
indeed the case with heavier targets \cite{BFG97,BFG95}.
However, since our main concern here was to use atomic potentials
at (very low) positive energies where $d$ and higher partial waves
contribute very little in light nuclei, we considered this procedure
acceptable. Note that we are using the same potential for all the
partial waves in the positive energy regime.

Very good fits to the data with reasonable values for the complex parameter
$b_0$ could be obtained only when a `finite range' was introduced in the
form of a Gaussian folding of a $\bar p N$ interaction into the nuclear
density distributions. This is in contrast to the case of heavier targets
where such a procedure was not necessary. Nevertheless, we consider the
resulting potential, with $b_0$=0.49+$i$3.0 fm and a Gaussian with a range
parameter of 1.4 fm folded into the nuclear density distribution
(hereafter referred to as potential (a)), as quite
reasonable. It fits within the errors the measured strong interaction
shifts and widths of the 2$p$ state in antiprotonic $^3$He and $^4$He.

\section{Results and discussion}
\label{sec:res}
Turning to positive energies, we first calculate the total reaction
(annihilation) cross section for 57 MeV/c $\bar p$ on $^4$He, 
in order to make it possible to compare results for He and Ne \cite{Bia00a}.
Using the above potential (a), the calculated
cross section is 901 mb, in excellent agreement with the 
interpolated value of 915$\pm$39 mb \cite{Bia00a}.
For  incident momenta of 47.0 and 70.4 MeV/c we calculate values of 1060 and
771 mb, respectively, where the experimental values \cite{Zen99b} are
979$\pm$145 and 827$\pm$38 mb, respectively. Reasonably good agreement
is therefore established between experiment and predictions made with
a potential derived from fits to $\bar p$ atoms of $^{3,4}$He.

Figure \ref{fig:sat} demonstrates the extreme strong-absorption
conditions which are relevant to the $\bar p$ nucleus interaction
at very low energies (and for $\bar p$ atoms). It shows calculated
reaction cross sections for $\bar p$ at 57 MeV/c on $^4$He
and Ne as function of the strength Im $b_0$ of the imaginary part
of the potential (a) described above, with the rightmost edge corresponding
to its nominal value Im $b_0$=3.0 fm. It is seen that as long as the
absorptivity (Im $b_0$) is very weak, less than 1\% of its nominal value,
$\sigma_R$ is approximately linear in Im $b_0$, which according to
Eq. (\ref{equ:sigma}) means that the $\bar p$ wavefunction depends
weakly on Im $b_0$. However, already  at below 5\% of the nominal value of
Im $b_0$ the reaction cross sections begin to saturate, much the same
as for the widths of deeply bound $\bar p$ atomic states \cite{FGa99a,FGa99b}.
The mechanism is the same in both cases, namely exclusion of the wavefunction
from the nucleus due to the absorption, which reduces dramatically the
overlap with the imaginary potential and consequently reduces  the integrals
in Eqs. (\ref{equ:gamma},\ref{equ:sigma}).
The onset of saturation is determined approximately by the strength parameter
2$\mu$(Im $V_{\rm opt}$)$R^2$, where $R$ is the radius of the nucleus.
Thus, saturation of $\sigma_R$ in Ne starts at a smaller
absorptivity than its onset in $^4$He. The Ne/$^4$He ratio of $\sigma_R$
values changes, due to this effect, from about 15 in the perturbative
regime of very weak absorptivity to about 3 in the strong-absorption regime.
However, the precise, detailed pattern of the change also depends on the real
part of the optical potential which may increase or reduce the exclusion of
the $\bar p$ wavefunction from the nucleus.

Figure \ref{fig:res} shows calculated $\bar p$
reaction cross sections at 57 MeV/c across the
periodic table. The dot-dashed line is for the above mentioned
potential (a) which is expected to be valid only in the immediate vicinity
of He. The dashed line (b) is for the first potential from Table 7 of
Ref. \cite{BFG97} which fits $\bar p$ atom data over the whole periodic
table, starting with carbon. This potential is not expected to
fit data for very light nuclei, and indeed it does not fit the $^4$He
annihilation cross section. However, it is noteworthy that for $A > $ 20
the two potentials predict almost the same cross sections and certainly
display a very smooth and similar dependence on $A$.
A smooth $A$ dependence of the calculated cross sections is generally
expected once several partial waves contribute, as we elaborate below.
Also shown in the figure is the recent experimental result \cite{Bia00a}
for Ne, with very limited accuracy.
Furthermore, for $A > $ 20 the points along the solid line are the
calculated $\bar n$ - nucleus total reaction cross sections, obtained
from potential (b) by switching off the Coulomb interaction.
The solid line is a fit to an $A^{1/3}$ power law which
appears to be appropriate to strong absorption of uncharged particles
at these very low energies, in contrast to the prediction of approximate
$A$ independence made in Ref. \cite{GKo97} for $\bar n$ annihilation
at ultra low energies where {\it only} $s$ waves contribute.
We checked that by reducing the absorptivity by 3 - 4 orders
of magnitude, a proportionality of $\sigma_R$ to $A$
as expected for weak absorption, is indeed observed.
Comparing the dashed line for
negatively charged particles with the solid line for uncharged particles,
it is clear that the $\sigma_R$ values obtained by
including the Coulomb interaction are considerably enhanced with respect
to the ones obtained without it. Attempting to fit the enhancement by a
$Z^{\alpha}$ power law, we find that $\alpha$ varies roughly between 1/3
for Ne to 1/2 for Pb. Attempts to identify more precisely this enhancement
across the periodic table have been recently addressed
in Ref. \cite{Bia00c}.

Finally, we discuss the $ZA^{1/3}$ scaling considered in
Refs. \cite{Bia00a,Bia00b} and in references quoted therein.
We note that an attractive Coulomb potential causes focussing of
partial-wave trajectories onto the nucleus, an effect which at very low
energies may be evaluated semiclassically (see Ref. \cite{Schiff}
for the applicability of the semiclassical approximation to Coulomb
scattering). The maximal orbital angular momentum $l_{\rm max}$,
for which the $\bar p$-nucleus distance of closest approach is smaller
than the nuclear radius $R$, at sufficiently low energy is given by

\begin{equation} \label{equ:lmax}
(l_{{\rm max}} + 1/2)^2  \approx  2 \eta k R \quad,
\end{equation}
where $k$ is the c.m. momentum and

\begin{equation} \label{equ:eta}
\eta = 2 Z / k_{L} a_{\rm B}^{(p)}
\end{equation}
is the corresponding Coulomb parameter.
Here $k_{L} = p_{L} / \hbar$, where $p_L$ is the
$\bar p$ laboratory momentum, and $a_{\rm B}^{(p)}$ is the Bohr
radius of the ${\bar p}p$ atom. Since 2$\eta >> kR$ at very low
energies and for high values of $Z$, the value of $l_{\rm max}$
greatly exceeds the value $kR$ which is appropriate to uncharged
particles ($\eta = 0$). For example, on Pb at $p_{L}=57$ MeV/c,
$l_{\rm max}$ is about 6 for $\bar p$, but is less than 2 for
$\bar n$. Therefore, for strong absorption, where all the lower
partial waves are totally absorbed, one obtains

\begin{equation} \label{equ:semi}
\sigma_R \approx \frac{\pi}{k^2} \sum_{l=0}^{l_{\rm {max}}}{(2l+1)}
\approx \frac{2\eta}{kR} \pi R^2 >> \pi R^2 \quad,
\end{equation}
which we rewrite as

\begin{equation} \label{equ:scaling}
\sigma_R \approx (4 \pi r_0 /a_{\rm B}^{(p)})
(Z A^{1/3} / k k_L)\quad,
\end{equation}
with $r_0$ defined by $R = r_{0} A^{1/3}$.
A similar approach was used by F\"aldt and Pilkuhn \cite{FPi72} to
calculate Coulomb corrections to $\pi^{\pm}$-nucleus total
cross sections near the 3,3 resonance. In the present case of very
low energies their correction becomes the dominant term, and the
resulting reaction cross section on the r.h.s. of Eq. (\ref{equ:semi})
is substantially higher than the black-disk value $\sigma _R = \pi R^2$
which for uncharged projectiles may be perceived as the unitarity limit
of $\sigma _R$. Indeed, the calculated saturation plateau values of 
$\sigma _R$ in Fig. \ref{fig:sat} are several times larger than 
the appropriate black-disk values.

In Figure \ref{fig:scale} we plot, for potential (b),
ratios of calculated total reaction cross sections
at 57 MeV/c to the scaling parameter $S=ZA^{1/3}/kk_L$.
It is clear that $\sigma_R$ does not scale with $S$ across the
periodic table. Also plotted is the asymptotic value

\begin{equation} \label{equ:asympt}
\sigma_R / S \to 4 \pi r_0 / a_{\rm B}^{(p)}
\approx 0.273 \quad,
\end{equation}
with $r_{0} = 1.25$ fm and $a_{\rm B}^{(p)} = 57.6$ fm.
For a given $\bar p$ beam energy this asymptotic value,
arising from Coulomb focussing,
is approached very slowly upon increasing $Z$.

Before concluding we briefly mention other derivations of the
$ZA^{1/3}$ dependence of $\sigma_R$. The proportionality to $Z$
is usually derived from the well-known Coulomb enhancement
Gamow factor which at low energies is well approximated by
$2 \pi \eta$ \cite{Bia00c}. However, at these very low energies,
the Coulomb wavefunctions are far from being constant over the
nuclear volume, so that the applicability of the Gamow factor
appears questionable.
The origin of a possible proportionality  of $\sigma_R$ to
$A^{1/3}$ is even less clear: for $s$-wave dominated $\bar p$
annihilation, as considered in
Refs. \cite{Bia00a,Bia00b}, it is due to the imaginary part
of the $\bar p$ scattering length which is assumed to closely
follow the nuclear radius $R$. However, for $\bar p$ strongly
absorptive potentials this imaginary part has invariably been
found to be about 1 fm across the periodic table \cite{Bat83}
(see also Ref. \cite{PBL00}), considerably less than $R$.
It is inconceivable that a proportionality to $R$ can arise
unless several partial waves contribute significantly.

\section{Conclusions}
\label{sec:conc}
\begin{figure}
\epsfig{file=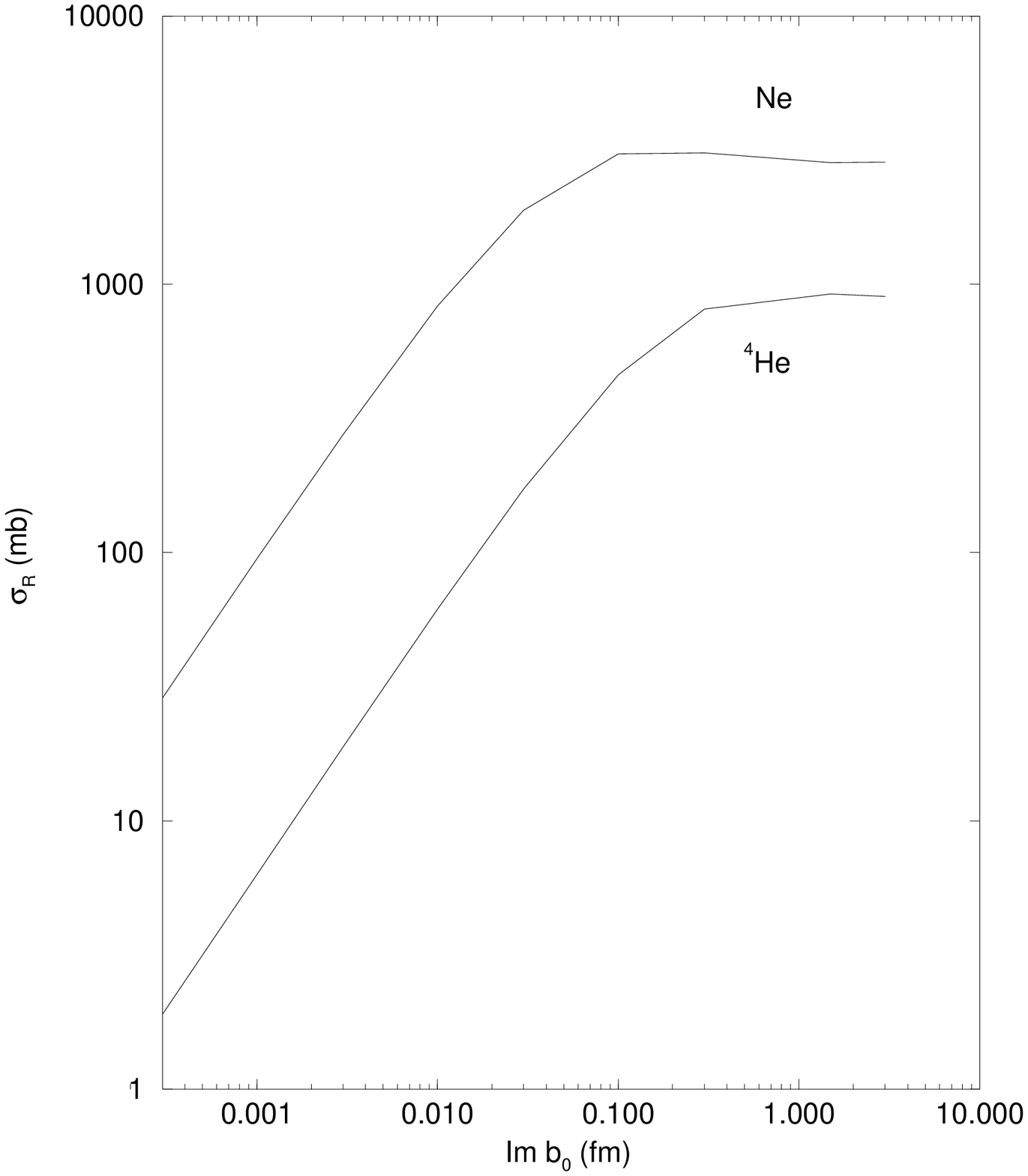,height=160mm,width=135mm,
bbllx=22,bblly=89,bburx=530,bbury=670}
\caption{Calculated total reaction cross sections for
57 MeV/c $\bar p$ on $^4$He and Ne as function
of the strength parameter Im $b_0$ of the optical potential (a).}
\label{fig:sat}
\end{figure}

\begin{figure}
\epsfig{file=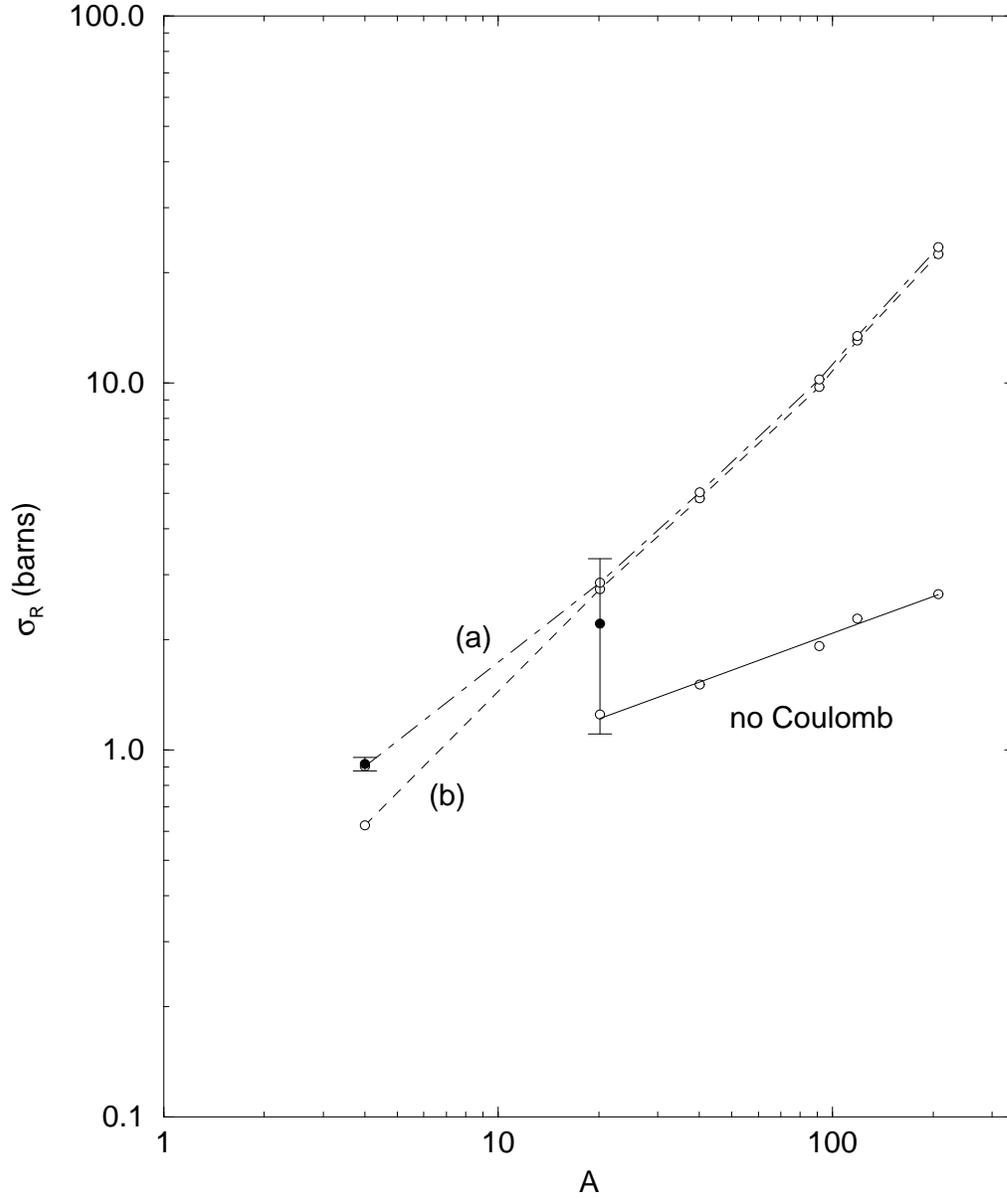,height=160mm,width=135mm,
bbllx=28,bblly=92,bburx=509,bbury=670}
\caption{Calculated $\bar p$ total
reaction cross sections (open circles) at 57 MeV/c
for potentials (a) and (b),
and for potential (b) but without the Coulomb interaction. Also shown are
the two data points for $^4$He and Ne.}
\label{fig:res}
\end{figure}

\begin{figure}
\epsfig{file=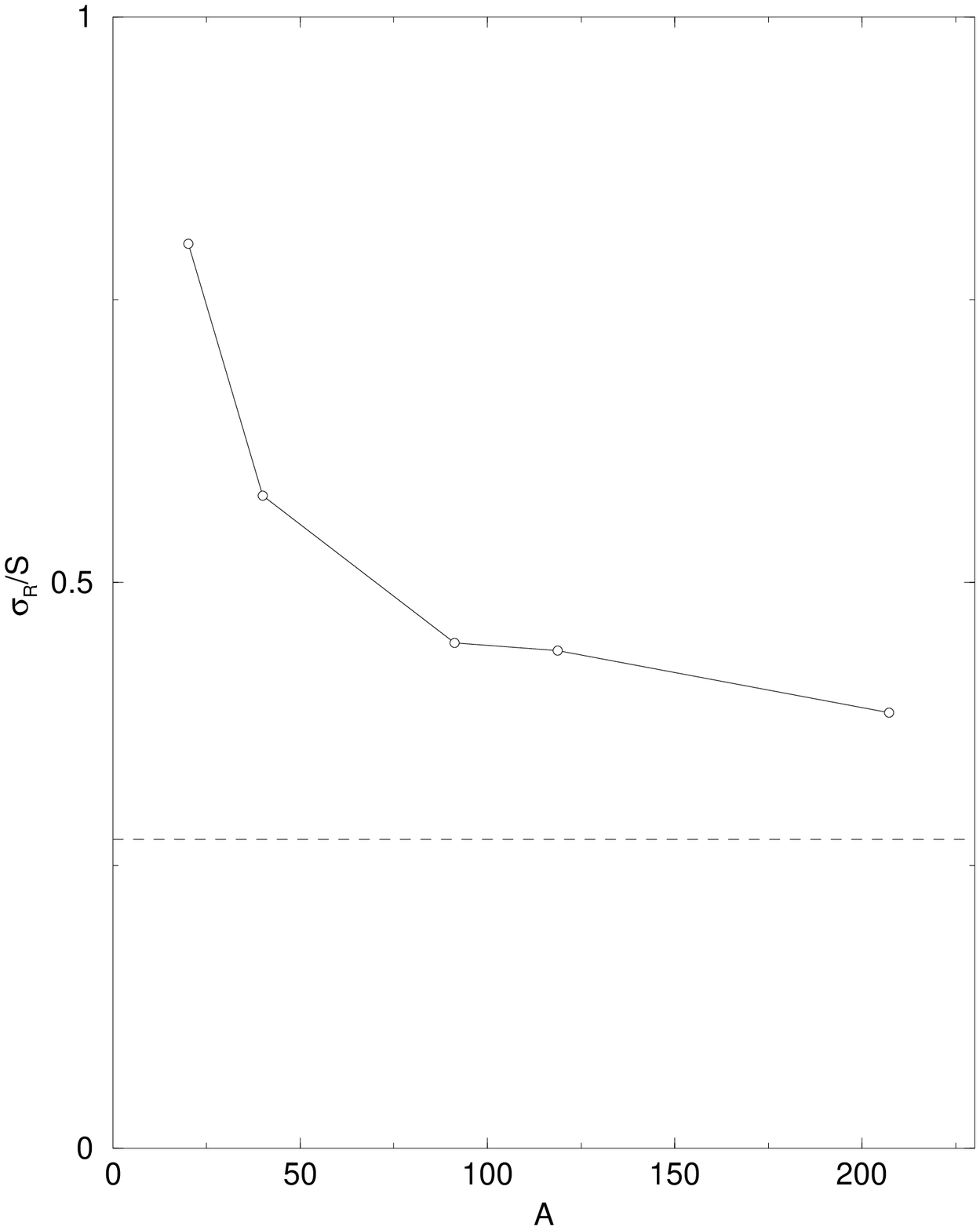,height=160mm,width=135mm,
bbllx=50,bblly=92,bburx=509,bbury=670}
\caption{Ratios of calculated
$\bar p$ total reaction cross sections at 57 MeV/c, for potential (b),
to the scale function S.
The dashed straight line provides the asymptotic value expected
for $\sigma_{R}/S$, see text.}
\label{fig:scale}
\end{figure}

In conclusion, we have shown that the recently reported annihilation
cross sections for $\bar p$ on $^4$He and Ne at 57 MeV/c are reproduced
very well by optical potentials which fit $\bar p$ atomic data in the
respective mass ranges (Fig. \ref{fig:res}). For these strongly
absorptive potentials, the asymptotic $ZA^{1/3}$ dependence
of the total reaction cross section was derived
from the Coulomb focussing effect at very low energies, but at 57 MeV/c
it was found not to be satisfied across the periodic table except
for unrealistically high values of $A$ (Fig. \ref{fig:scale}).
The correct dependence is determined by the
saturation property of total reaction cross sections at very low
energies (Fig. \ref{fig:sat}). The apparent suppression of $\bar p$
total annihilation cross sections measured at $p_L < 100$ MeV/c
is a direct consequence of the strong absorption, and in particular its
saturation property. Predictions have been given for $\bar p$ total
annihilation cross sections at 57 MeV/c across the periodic table
which would complement and enrich the information deduced from the
measurements already available for $\bar p$ atoms.

A more extended discussion of optical model calculations and their
implications for $\bar p$ atoms and $\bar p$ annihilation on nuclei,
including a connection to $\bar p p$ annihilation at $p_{L}<200$
MeV/c, is included in a forthcoming publication \cite{BFG00}.

\vspace{10mm}

CJB wishes to thank the Hebrew University for support for a visit
during which this work was started.
\newline
This research was partially supported by the Israel Science Foundation.

\begin{thebibliography}{9}
	\bibitem{Zen99a}  A. Zenoni et al., Phys. Lett. B 461 (1999) 405.

	\bibitem{Zen99b}  A. Zenoni et al., Phys. Lett. B 461 (1999) 413.

	\bibitem{Bia00a}  A. Bianconi et al., Phys. Lett. B 481 (2000) 194.

        \bibitem{Bia00b}  A. Bianconi, G. Bonomi, M.P. Bussa, E. Lodi Rizzini,
        L. Venturelli, A. Zenoni, Phys. Lett. B 483 (2000) 353.
	
        \bibitem{BFG97}  C.J. Batty, E. Friedman, A. Gal, Phys. Rep. 287
	(1997) 385.

        \bibitem{BFG95}  C.J. Batty, E. Friedman, A. Gal, Nucl. Phys. A
        592 (1995) 487.

        \bibitem{SMC79}  K. Stricker, H. McManus, J.A. Carr,
        Phys. Rev. C 19 (1979) 929.

	\bibitem{FGJ91}  E. Friedman et al., Phys. Lett. B 257 (1991) 17.

        \bibitem{PBL00}  K.V. Protasov, G. Bonomi, E. Lodi Rizzini,
        A. Zenoni, Eur. Phys. J. A 7 (2000) 429.

	\bibitem{FGa99a}  E. Friedman, A. Gal, Phys. Lett. B 459 (1999) 43.
	
        \bibitem{FGa99b}  E. Friedman, A. Gal, Nucl. Phys. A 658 (1999) 345.

        \bibitem{WGN85}  S. Wycech, A.M. Green, J.A. Niskanen, Phys. Lett.
        152 B (1985) 308.

        \bibitem{LTa90}  G.P. Latta, P.C. Tandy, Phys. Rev. C 42 (1990) R1207.

        \bibitem{KPV00}  V.A. Karmanov, K.V. Protasov, A. Yu. Voronin,
        Eur. Phys. J. A (2000), in press (nucl-th/0006041).

        \bibitem{Sch91}  M. Schneider et al., Z. Phys. A 338 (1991) 217.

        \bibitem{GKo97}  Ye.S. Golubeva, L.A. Kondratyuk, Nucl. Phys. B
        (Proc. Suppl.) 56A (1997) 103.

        \bibitem{Bia00c}  A. Bianconi, G. Bonomi, E. Lodi Rizzini,
        L. Venturelli, A. Zenoni, Phys. Rev. C 62 (2000) 014611.

        \bibitem{Schiff}  L.I. Schiff, Quantum Mechanics, third edition
        (McGraw Hill, New York, 1968)

        \bibitem{FPi72} G. F\"aldt, H. Pilkuhn, Phys. Lett. 40 B
        (1972) 613.

        \bibitem{Bat83}  C.J. Batty, Nucl. Phys. A 411 (1983) 399.

        \bibitem{BFG00}  C.J. Batty, E. Friedman, A. Gal, to be
        submitted to Nucl. Phys. A (2000).

\end{thebibliography}
\end{document}